\documentclass{INTERSPEECH2023}

\interspeechcameraready

\usepackage{multirow}
\usepackage{subfigure}
\usepackage{hyperref}
\usepackage{xcolor}
\usepackage{url}

\title{RMVPE: A Robust Model for Vocal Pitch Estimation in Polyphonic Music}
\name{Haojie Wei$^{1}$, Xueke Cao$^{2}$, Tangpeng Dan$^1$, Yueguo Chen$^{1,*}$~\thanks{$^{*}$ Corresponding Author}}
\address{
  $^1$School of Information, Renmin University of China, Beijing, China\\
  $^2$Institute of Information Science, Beijing Jiaotong University, Beijing, China}
\email{weihaojie@ruc.edu.cn, caoxueke19@163.com, tangpengdan@ruc.edu.cn, chenyueguo@ruc.edu.cn}

\begin{document}

\maketitle
 
\begin{abstract}
Vocal pitch is an important high-level feature in music audio processing. 
However, extracting vocal pitch in polyphonic music is more challenging due to the presence of accompaniment. 
To eliminate the influence of the accompaniment, most previous methods adopt music source separation models to obtain clean vocals from polyphonic music before predicting vocal pitches. 
As a result, the performance of vocal pitch estimation is affected by the music source separation models. 
To address this issue and directly extract vocal pitches from polyphonic music, we propose a robust model named RMVPE. 
This model can extract effective hidden features and accurately predict vocal pitches from polyphonic music. 
The experimental results demonstrate the superiority of RMVPE in terms of raw pitch accuracy (RPA) and raw chroma accuracy (RCA).
Additionally, experiments conducted with different types of noise show that RMVPE is robust across all signal-to-noise ratio (SNR) levels. The code of RMVPE is available at \url{https://github.com/Dream-High/RMVPE}.
\end{abstract}


\section{Introduction}\label{sec:intro}
Pitch Estimation (PE), also known as pitch tracking or fundamental frequency ($f0$) estimation, is important in music signal processing. 
It plays a crucial role in melody extraction, and $f0$ can reflect various features of music audio and speech~\cite{carroll2011fundamental}. As a result, it is widely used in music information retrieval (MIR)~\cite{mir} and speech analysis (SA)~\cite{sa}.
A specific case of pitch estimation is vocal pitch estimation, which is extracting vocal pitches from clean vocals. 
It should be noted that clean vocals are a type of monophonic music. 
While for extracting vocal pitches in polyphonic music, the vocal pitch estimation task becomes more challenging due to the presence of accompaniment.

The pitch estimation task in music information retrieval has been researched for decades. 
Many algorithms have been proposed to solve this problem, primarily falling into two categories: traditional heuristic methods and data-driven based methods.
Traditional methods, such as YIN~\cite{de2002yin}, SWIPE~\cite{swipe}, and pYIN~\cite{mauch2014pyin}, employ specific candidate-generating functions to produce pitch results. 
These methods can get a well pitch result from clean vocals or other monophonic music.
Data-driven based methods, including CREPE~\cite{kim2018crepe}, DeepF0~\cite{deepf0}, and HARMOF0~\cite{harmof0}, utilize supervised training of CNN-based models for pitch estimation. 
These methods outperform traditional methods when applied to clean vocals.
Additionally, data-driven based methods have the ability to process low-level noise as they can capture deep-level features that are more suitable for pitch estimation, even in low-noise environments.

The aforementioned methods belong to single-pitch estimation (SPE) since they are designed for monophonic music and predict no more than one pitch per frame. 
Consequently, their performance sharply decreases when extracting vocal pitches from polyphonic music due to the presence of accompaniment. However, the polyphonic music, which consists vocals and accompaniment (e.g., pop music), has a large proportion in music data.
It is more universal than clean vocals in music audio.
Moreover, vocal pitch is an important high-level feature in polyphonic music and is effective for music information retrieval. 
Therefore, extracting vocal pitches from polyphonic music is an important and challenging task in music information retrieval.
It should be noted that, the vocal pitch estimation task also belongs to SPE since it only extracts vocal pitches in polyphonic music, and each frame has at most one vocal pitch.

Some existing work has also tried to extract vocal pitches in polyphonic music using two main methods: the pipeline method and the end-to-end method.
The pipeline method uses existing models to extract vocal pitches in polyphonic music. 
Firstly, a music source separation model (e.g., Open-Unmix~\cite{open2019}, Demucs~\cite{demucs2019}, Spleeter~\cite{spleeter2020}, etc.) is applied to extract clean vocals. 
Then, a pitch estimation model (e.g., pYIN~\cite{mauch2014pyin}, CREPE~\cite{kim2018crepe}, HARMOF0~\cite{harmof0}, etc.) is used to extract vocal pitches based on the predicted vocals. 
It is evident that the performance of vocal pitch estimation is highly related to both music source separation models and pitch estimation models.
The end-to-end method designs a robust model to eliminate the influence of accompaniment. For example, CRN-Raw~\cite{crn2019} uses a deep convolutional residual network to directly extract vocal pitches from the raw waveform of polyphonic music. 
JDC~\cite{jdc2022} employs a joint model of singing voice detection and classification to process the accompaniment. 
However, neither of these methods can extract highly accurate vocal pitches in polyphonic music.


In this paper, we propose a \textbf{R}obust \textbf{M}odel for \textbf{V}ocal \textbf{P}itch \textbf{E}stimation (RMVPE) in polyphonic music. RMVPE is inspired by music source separation models, which can extract clean vocals from polyphonic music. Our model utilizes deep U-Net and GRU to directly extract vocal pitches from polyphonic music.
By this way, RMVPE can not only extract vocal pitches from polyphonic music, but also has robustness to different types of noise.
Additionally, RMVPE have the comparable performance on clean vocals or other monophonic music.

\section{Method} \label{sec:method}
The vocal pitch estimation task in this paper belongs to SPE. To solve this task, we propose a robust model called RMVPE.
The overall structure of RMVPE and the loss function are introduced in Section~\ref{sec:model} and Section~\ref{sec:loss}, respectively.

\vspace{-2mm}
\subsection{Model Architecture}\label{sec:model}
\begin{figure}[htp]
\vspace{-2mm}
\centerline{\includegraphics[width=\columnwidth]{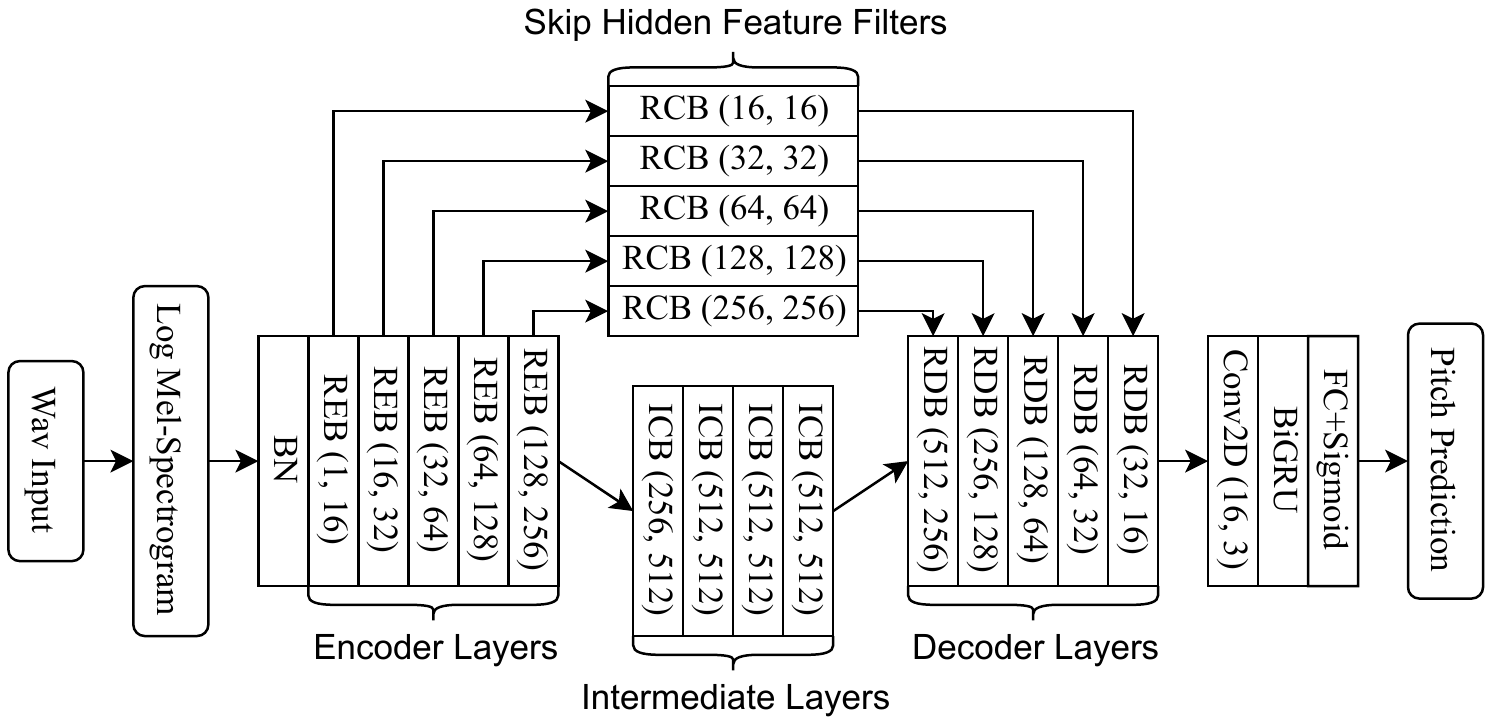}}
\vspace{-2mm}
\caption{The overall structure of our proposed model RMVPE.}
\label{fig:structure}
\vspace{-2mm}
\end{figure}

\begin{figure}[htp]
\vspace{-4mm}
    \centering
    \subfigure[]{
        \label{fig:2a}
        \centering
        \includegraphics[height=0.2232\columnwidth]{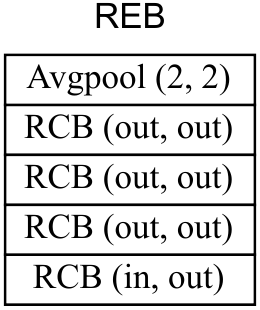}}
    \hspace{3mm}
    \subfigure[]{
        \label{fig:2c}
        \centering
        \includegraphics[height=0.3456\columnwidth]{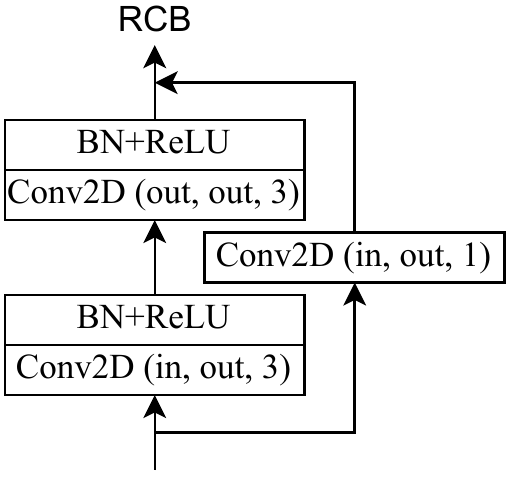}}
    \hspace{3mm}
    \subfigure[]{
        \label{fig:2b}
        \centering
        \includegraphics[height=0.2619\columnwidth]{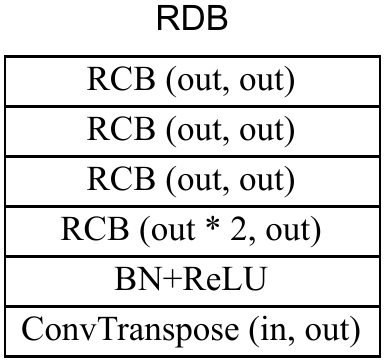}}
    \vspace{-2mm}
    \caption{(a) Residual Encoder Block (REB), (b) Residual Convolutional Block (RCB), (c) Residual Decoder Block (RDB)}
    \vspace{-2mm}
\end{figure}

We use the log mel-spectrograms $\text{X}_{T\times F}$ as the input features and the probability matrix ${Y}_{T\times 360}$ to represent the predicted pitches. In this way, the pitch estimation task can be formally written as, $\mathit{F}: X_{T\times F} \to Y_{T\times 360}$, where $T$ represents frames of the audio and $F$ is the logarithmically-spaced frequency bins of spectrogram. 
As shown in Figure~\ref{fig:structure}, 
RMVPE mainly consists of four components, encoder layers, skip hidden feature filters, intermediate layers and decoder layers. The details of each component are introduced as follows:

The input features firstly pass through a batch normalization (BN)~\cite{bn} layer. After that, there are encoder layers, which contain 5 residual encoder blocks (REBs). Each REB contains 4 residual convolutional blocks (RCBs) followed by an average pooling layer with a kernel size $2\times2$ to extract the hidden features as shown in Figure~\ref{fig:2a}. 
The details of RCB are shown in Figure~\ref{fig:2c}. 
There are two convolutional layers with $3\times3$ kernels. After each convolutional layer, there is a BN layer and a ReLU function. Additionally, there is a shortcut connection between the input and the output of a RCB.

After the encoder layers, there are intermediate layers and skip hidden feature filters. 
The intermediate layers take the output of the encoder layers as input and contain four intermediate convolutional blocks (ICBs). 
Each ICB has the same structure as an REB without the average pooling layer. 
The skip hidden feature filters consist of 5 RCBs and take the output of each REB before the average pooling layer as input. 
The output of each RCB in the skip hidden feature filters is used as the input for the corresponding RDB in the decoder layers.


The decoder layers are the final component of RMVPE, and they consist of 5 residual decoder blocks (RDBs). 
Each RDB includes a transposed convolutional layer with a kernel size of $3\times3$ and stride of $2\times2$ to upsample feature maps, followed by 4 RCBs, as shown in Figure~\ref{fig:2b}. After the decoder layers, a convolutional layer with a kernel size of $3\times3$, a bidirectional GRU \cite{gru} with 256 units, and a fully connected sigmoid layer with 360 outputs are applied.

\subsection{The Loss Function}\label{sec:loss}
Each frame of the output $Y_{T\times 360}$ is a 360-dimensional vector $y$, which represents the logarithmic scale pitches measured in terms of cents. The cent is a logarithmic unit of measure used for musical intervals relative to a reference pitch $f_{\textit{ref}}$ in Hz, defined as a function of frequency $f$ in Hz:
\begin{equation} 
\label{eq:f2c}
c(f) = 1200\times\log_2{\frac{f}{f_{\textit{ref}}}}
\end{equation}
where we set $f_{\textit{ref}}$ as 10 Hz here followed by~\cite{kim2018crepe}. The output vector $y$ corresponds to the frequency bin, which ranges from $C1$ (32.7 Hz) to $B6$ (1975.5 Hz) with 20 cents of intervals. By this way, the output vector $y$ fully covers the range of most melodic instruments including the singing voice.

When calculating the pitch value from $\hat{y}$, we use the local weighted average of pitches closest to the frequency bins with the highest peak value, as shown in Eq.~\ref{eq:cent}:
\begin{equation} 
\label{eq:cent}
\hat{c} = \sum_{m-4}^{m+4} (\hat{y_{i}} c_i) \bigg/ \sum_{m-4}^{m+4} (\hat{y_{i}}) \quad m=\mathop{\arg\max}\limits_{i} \hat{y_{i}}
\end{equation}
Then the predicted frequency is calculated by Eq.~\ref{eq:c2f}. 
The \textit{confi} in Eq.~\ref{eq:c2f} represents the voicing confidence, which reflects the confidence in the presence of a vocal pitch.
In this paper, we set the default threshold (\textit{th}) to 0.5.
\begin{equation} 
\label{eq:c2f}
\hat{f}= 
\begin{cases}
    f_{\textit{ref}}\times 2^{\hat{c}/1200}& {\textit{confi} \geq \textit{th}}\\
    0&{\textit{confi} < \textit{th}}
\end{cases}
\quad \textit{confi}=\mathop{\max}\limits_{i} \hat{y_{i}}
\end{equation}

When training the model, we transform the ground-truth pitch of each frame to a 360-dimensional one-hot vector $y$. To address the class imbalance between positive and negative categories, we use the weighted cross entropy loss. The loss function is defined as follows:
\begin{equation} 
\mathcal{L}(y, \hat{y}) = -\sum_{i}^{360} (\omega y_{i} \log \hat{y_{i}}+(1-y_{i}) \log(1- \hat{y_{i}}))
\end{equation}
where we set $\omega$ as 5, as it achieved the best performance in this paper. 

\section{Experiment} \label{sec:experiment}
In this section, we firstly evaluate our model on three public polyphonic music datasets to show our model is practical in real music scenarios. Then we evaluate RMVPE with different types of noise to show our model is robust to noise. Finally, the experimental results show that RMVPE is also general with monophonic music since it has the comparable performance on clean vocals or other monophonic music. 

\subsection{Datasets and Evaluation Metrics}
To compare with previous models fairly, we use four public datasets MDB-stem-synth~\cite{salamon2017analysis}, MIR-1K~\cite{mir1k}, Cmedia
\footnote{\url{https://www.music-ir.org/mirex/wiki/2020:Singing\_Transcription\_from\_Polyphonic\_Music}} 
and MIR\_ST500~\cite{mirst500}. We use raw pitch accuracy (RPA), raw chroma accuracy (RCA) and overall accuracy (OA), which are computed by mir\_eval\cite{raffel2014mir_eval} library to evaluate different models.

\textbf{Datasets.} \textit{MDB-stem-synth} is a monophonic music dataset for pitch estimation, which contains 230 resynthesized music files corresponding perfect $f0$ annotation. 
\textit{MIR-1K} is a dataset designed for singing voice separation, which has accompaniment on right channels and clean vocals on left channel. It also has pitch labels of vocal parts.
\textit{MIR\_ST500} and \textit{Cmedia} are polyphonic music datasets, which have the YouTube URL of pop music and the pitch label of vocal parts.

\textbf{Evaluation Metrics.} Raw pitch accuracy (RPA) computes the proportion of melody frames in the reference for which the predicted pitch is within $\pm50$ cents of the ground truth pitch.
While raw chroma accuracy (RCA) measures the raw pitch accuracy ignoring the octave errors. Overall accuracy (OA) computes the proportion of all frames correctly estimated by the algorithm, including whether non-melody frames where labeled by the algorithm as non-melody.

\subsection{Experimental setup}
The music audio is sampled at 16 kHz and then transformed into mel-spectrograms with log amplitude which has 256 mel bins. The hop length of mel-spectrograms is 320 (20ms) and the Hann window size is 2048. We cut the frequency between 30Hz and 8000Hz to extract the mel-spectrograms. librosa~\cite{mcfee2015librosa} is adopted to finish the above audio processing.

Our model is trained using Adam optimizer~\cite{kingma2014adam}. The learning rate is initialized as 0.0005 and reduced by 0.98 of the previous learning rate every 10 epochs and the batch size is set as 16.
For MDB-stem-synth, MIR-1K and Cmedia datasets, we randomly spilt these datasets (at song-level) into training (80\%) set and testing (20\%) set.
The spilt way of MIR\_ST500 is introduced in \cite{mirst500}. For all datasets, we split the audio recordings into 2.56-second segments when training the model. 

\subsection{Results on Polyphonic Music}
We firstly compare RMVPE with several previous models on polyphonic music datasets MIR-1K, MIR\_ST500 and Cmedia. For PYIN~\cite{mauch2014pyin}, CREPE~\cite{kim2018crepe} and HARMOF0~\cite{harmof0}, they are proposed for monophonic music, thus we firstly get clean vocals by Spleeter~\cite{spleeter2020} and evaluate these models based on predicted vocals. The above methods belong to the pipeline method as illustrated in Section~\ref{sec:intro}. 
For JDC~\cite{jdc2022} and CRN-Raw~\cite{crn2019}, they belong to the end-to-end method. We use the open checkpoint to evaluate JDC, since JDC is a joint model, which can process the accompaniment. The CRN-Raw model is retrained by using the same datasets as the RMVPE.
\begin{table}[htbp]
\centering
\caption{Average RPA(\%), RCA(\%) and OA(\%) on three different polyphonic music datasets.}
\label{tab:mixture}
\resizebox{\columnwidth}{!}{
\begin{tabular}{c|c|ccc}
\hline
\multirow{2}{*}{Methods} & \multirow{2}{*}{Metrics} & \multicolumn{3}{c}{Datasets}                      \\ \cline{3-5} 
                         &                          & MIR-1K & MIR\_ST500     & Cmedia         \\ \hline
\multirow{3}{*}{PYIN~\cite{mauch2014pyin}}    & RPA                  & 77.29±9.05           & 63.85±8.49          & 59.69±9.17          \\ 
                         & RCA                  & 77.86±9.52           & 65.01±8.14          & 60.92±9.83          \\ 
                          & OA                   & 71.79±9.97           & 65.27±7.66          & 60.25±8.73          \\ \hline
\multirow{3}{*}{CREPE~\cite{kim2018crepe}}   & RPA                  & 91.05±8.46           & 81.61±8.03               & 74.61±5.92               \\ 
                         & RCA                  & 92.16±7.30           & 82.36±7.70               & 75.05±5.57               \\
                         & OA                   & 88.71±7.44           & 76.59±6.35          & 75.19±5.23          \\ \hline
\multirow{3}{*}{HARMOF0~\cite{harmof0}} & RPA                  & 88.63±8.12           & 83.00±7.47         & 79.08±6.37          \\ 
                         & RCA                  & 88.97±7.61           & 83.35±7.09          & 79.64±5.95          \\ 
                         & OA                   & 87.64±7.55           & 81.80±5.94          & 81.83±5.56          \\\hline
\multirow{3}{*}{CRN-Raw~\cite{crn2019}} & RPA                  & 82.75±9.56           & 76.64±8.52         & 75.88±8.06          \\ 
                         & RCA                  & 91.92±9.39           & 82.48±7.96          & 79.86±7.23          \\ 
                         & OA                   & 85.53±7.99           & 80.81±8.09         & 78.85±7.85          \\\hline
\multirow{3}{*}{JDC~\cite{jdc2022}}     & RPA                  & 82.28±7.86           & 80.09±9.24          & 79.00±7.83          \\ 
                         & RCA                  & 82.98±7.57           & 80.38±9.08          & 79.25±7.57          \\ 
                         & OA                   & 78.61±6.84           & 82.86±5.72          & 82.64±6.01          \\ \hline
\multirow{3}{*}{$\text{RMVPE}_\textit{Poly}$}    & RPA                  & \textbf{95.42±3.97}  & \textbf{89.32±5.64} & \textbf{83.57±5.62} \\ 
                         & RCA                  & \textbf{95.84±3.51}  & \textbf{89.84±5.11} & \textbf{84.04±5.28} \\ 
                         & OA                   & \textbf{91.86±5.08}           & \textbf{84.54±5.17}          & \textbf{85.09±5.01}          \\ \hline
\end{tabular}
}
\vspace{-2mm}
\end{table}

As shown in Table~\ref{tab:mixture}, we find that our model has the best performance at RPA, RCA and OA compared with other models. Specifically, on MIR\_ST500 dataset, $\text{RMVPE}_\textit{Poly}$ outperforms the second best model HARMOF0 by 6.32\% at RPA, by 6.49\% at RCA, by 2.74\% at OA. This is because the encoder layers and decoder layers can directly extract effective hidden features from polyphonic music for vocal pitch estimation. It should be noted that $\text{RMVPE}_\textit{Poly}$ means the model is trained on polyphonic music data. The above results show that our model is effective in real music scenarios since it achieves state-of-the-art performance on different polyphonic music datasets.

\begin{figure*}[htp]
\centerline{\includegraphics[width=0.9\textwidth]{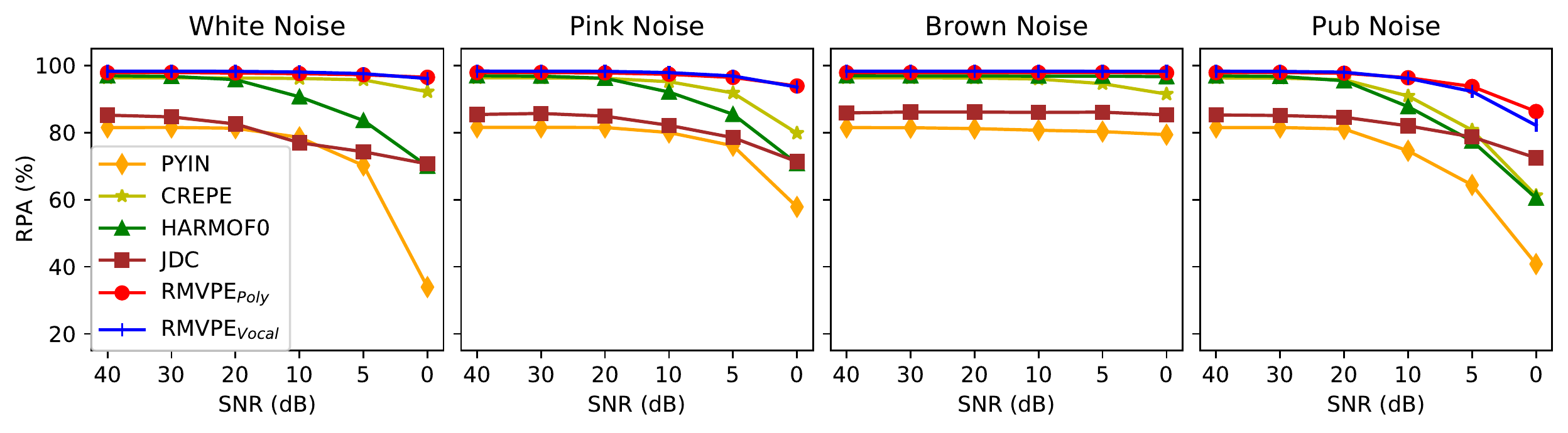}}
\vspace{-2mm}
\caption{The RPA (\%) performance under different noise levels.}
\vspace{-4mm}
\label{fig:noise}
\end{figure*}

\subsection{Results with Different Types of Noise}\label{sec:noise}
Our model can not only predict vocal pitches from polyphonic music, but also extract vocal pitches with different types of noise. To verify RMVPE is more robust with different types of noise than previous models,
we use the clean vocals in MIR-1K~\cite{mir1k} adding different types of noise to evaluate previous models and ours.
Specifically, we add different types of noise by using Audio Degradation Toolbox (ADT)~\cite{adt}. 
In this paper, we use four types of noise provided by ADT, white noise, pink noise, brown noise and pub noise.
The white noise is a random signal which has constant power over all frequencies and the pink noise is a signal with a power density which decreases with increasing frequency. 
The brown noise is similar to pink noise, it has the highest power spectral density in low frequencies.
The pub noise is a real recording of the sound in a crowded pub. We set the signal-to-noise ratio (SNR) value as 0 dB in this experiment.

We retrain CREPE~\cite{kim2018crepe} and HARMOF0\cite{harmof0} using clean vocals of MIR-1K here. 
$\text{RMVPE}_\textit{Vocal}$ and $\text{RMVPE}_\textit{Poly}$ have the same model structure, but they are trained on different types of music datasets. $\text{RMVPE}_\textit{Poly}$ is trained on mixture (polyphonic) music of MIR-1K and $\text{RMVPE}_\textit{Vocal}$ is trained on clean vocals of MIR-1K. The results of GIO~\cite{GIO2022} are copied from the original paper, because our experimental settings are the same as that in GIO.
All results on vocals with different types of noise are summarized in Table~\ref{tab:noise}.

\begin{table}[htb]
\centering
\caption{Comparison with previous work with different types of noise when signal-to-noise ratio (SNR) is 0 dB on MIR-1K dataset (\%).}
\label{tab:noise}

\resizebox{\columnwidth}{!}{
\begin{tabular}{c|c|cccc|c}
\hline
           Methods                     & Metrics    & white & pink  & brown & pub   & Avg. \\ \hline
\multirow{2}{*}{PYIN~\cite{mauch2014pyin}}           & RPA & 33.94 & 57.83 & 79.36 & 40.80 &52.98      \\ 
                                & RCA & 34.21 & 58.30 & 79.80 & 48.64 &55.24      \\ \hline
\multirow{2}{*}{CREPE~\cite{kim2018crepe}}          & RPA & 92.15 & 89.01 & 91.47 & 61.22 &  81.16    \\ 
                                & RCA & 95.61 & 90.07 & 93.75 & 71.24 &87.67      \\ \hline
\multirow{2}{*}{HARMOF0~\cite{harmof0}}        & RPA & 74.61 & 74.85 & 97.18 & 61.62 &77.07      \\ 
                                & RCA & 75.18 & 76.41 & 97.24 & 65.05 &78.48      \\ \hline
\multirow{2}{*}{JDC~\cite{jdc2022}}            & RPA & 70.65 & 71.32 & 85.25 & 72.44 &74.92      \\  
                                & RCA & 72.01 & 82.24 & 85.45 & 73.14 &77.96      \\ \hline
\multirow{2}{*}{GIO~\cite{GIO2022}}            & RPA & 94.20 & 89.10 & 97.10 & 68.60 &87.25      \\ 
                                & RCA & 94.90 & 91.40 & 97.50 & 74.70 &89.63      \\ \hline
\multirow{2}{*}{$\text{RMVPE}_\textit{Poly}$} & RPA & \textbf{96.47} & \textbf{93.87} & 97.78 & \textbf{86.26} &\textbf{93.60}      \\ 
                                & RCA & \textbf{96.96} & \textbf{95.13} & 97.95 & \textbf{87.44} &\textbf{94.37}      \\ \hline
\multirow{2}{*}{$\text{RMVPE}_\textit{Vocal}$}   & RPA & 96.09 & 93.55 & \textbf{98.25} & 82.13 &92.51      \\ 
                                & RCA & 96.33 & 95.02 & \textbf{98.30} & 84.39 &93.51      \\ \hline
\end{tabular}
}
\end{table}

For $\text{RMVPE}_\textit{Poly}$ and $\text{RMVPE}_\textit{Vocal}$, the performance of RPA and RCA is better than the previous models. 
With the white, pink and pub noise, $\text{RMVPE}_\textit{Poly}$ performs better than $\text{RMVPE}_\textit{Vocal}$. Especially for pub noise, $\text{RMVPE}_\textit{Poly}$ gets an improvement of 4.13\% at RPA and 3.05\% at RCA compared with $\text{RMVPE}_\textit{Vocal}$. This is because pub is a more complex noise than others and RMVPE trained on mixture music is more robust to noise. While for brown noise, $\text{RMVPE}_\textit{Vocal}$ outperforms $\text{RMVPE}_\textit{Poly}$ by 0.47\% at RPA, by 0.35\% at RCA. This is because brown noise has most of its energy at low frequency and affects less to the pitch estimation results. These results show that RMVPE is stable at different training conditions. Thus, RMVPE is easier to use in practice than other models because training our model requires music data with less limitation (both polyphonic music and clean vocals are OK).

After that, Table~\ref{tab:noise} also shows that RMVPE achieves the highest performance with all types of noise compared with previous models. With pink noise, $\text{RMVPE}_\textit{Poly}$ outperforms the second best model GIO by 4.77\% at RPA, by 3.73\% at RCA. With pub noise, $\text{RMVPE}_\textit{Poly}$ outperforms the second best model GIO by 17.66\% at RPA, by 12.74\% at RCA. And there is an average improvement of 6.35\% at RPA, 4.74\% at RCA. The RMVPE improves significantly at different types of noise, especially with the pub noise. These results show that our model is more robust with different types of noise.

\subsection{Noise Robustness}
In order to further evaluate the performance of different models with different levels of noise. We add the noise to clean vocals with different signal-to-noise ratio (SNRs). The details of different types of noise and the settings of previous models are the same as Section~\ref{sec:noise}. In this experiment, we use six different SNR values: 40, 30, 20, 10, 5 and 0 dB. 

In Figure~\ref{fig:noise}, we find that $\text{RMVPE}_\textit{Poly}$ maintains the highest performance at RPA, and the performance of $\text{RMVPE}_\textit{Vocal}$ is almost the same as $\text{RMVPE}_\textit{Poly}$. This is because both of them can extract effective hidden features for vocal pitch estimation and $\text{RMVPE}_\textit{Poly}$ trained on polyphonic music is more robust to noise.
It is obviously that the performance of different models decreases sharply when SNR is under 10 dB. Especially on pub noise, since the pub noise is a complex environmental noise. Brown noise is the exception where the performance decreases less, this is because brown noise has most of its energy at low frequencies. 
Moreover, our model decreases less than other models at all types of noise as shown in Figure~\ref{fig:noise}. The second best model CREPE decreases 3.96\% under white noise, 15.34\% under pink noise, 4.48\% under brown noise and 29.67\% under pub noise from 10 dB to 0 dB. While $\text{RMVPE}_\textit{Poly}$ decreases 1.08\% under white noise, 3.48\% under pink noise, 0.08\% under brown noise and 10.07\% under pub noise from 10 dB to 0 dB. These results show that RMVPE has the ability to process different types of noise at all levels of noise.

\subsection{Results on Clean Vocals}
We also evaluate the performance of RMVPE on monophonic music (MDB-stem-synth~\cite{salamon2017analysis}) and clean vocals (MIR-1K~\cite{mir1k}). The compared algorithms are PYIN~\cite{mauch2014pyin}, CREPE~\cite{kim2018crepe}, HARMOF0~\cite{harmof0} and JDC~\cite{jdc2022}. Different from experiments in Section~\ref{sec:noise}, CREPE and HARMOF0 is retrained on MDB-stem-synth and clean vocals of MIR-1K. 

\begin{table}[htb]
\centering
\caption{Average RPA(\%), RCA(\%) and OA(\%) on clean vocals or monophonic music datasets.}
\label{tab:clean}
\resizebox{\columnwidth}{!}{
\begin{tabular}{c|c|ccc}
\hline
\multirow{2}{*}{Datasets} & \multirow{2}{*}{Methods} & \multicolumn{3}{c}{Metrics}                      \\ \cline{3-5} 
                         &                          & RPA & RCA     & OA         \\ \hline
\multirow{5}{*}{MDB-stem-synth}        &PYIN~\cite{mauch2014pyin} & 65.83±21.29            & 67.01±21.36           & 77.57±16.33          \\ 
                         &CREPE~\cite{kim2018crepe}& 97.50±4.25            & 97.97±3.22      &98.41±2.93              \\ 
                         &HARMOF0~\cite{harmof0}& \textbf{97.94±2.19}            & \textbf{98.02±2.12}      & \textbf{98.47±1.76}         \\ 
                         &JDC~\cite{jdc2022}& 62.61±26.43 & 62.81±26.50  &68.03±30.69         \\ 
                         &$\text{RMVPE}_\textit{Vocal}$  & 97.11±2.70            & 97.12±2.69 &97.68±2.09 \\ \hline
\multirow{6}{*}{MIR-1K (vocals)}        &PYIN~\cite{mauch2014pyin} &74.71±10.37&74.99±9.48 &79.07±7.31        \\ 
                         &CREPE~\cite{kim2018crepe}& 95.66±4.07            & 96.52±2.87  &95.56±2.90                \\ 
                         &HARMOF0~\cite{harmof0}& 96.07±3.54            & 96.58±2.84 &96.31±2.43            \\ 
                         &JDC~\cite{jdc2022}&68.96±10.78&69.54±10.56&  66.85±9.86        \\ 
                         &$\text{RMVPE}_\textit{Poly}$ &96.71±3.62 &96.80±3.23 &95.68±2.40     \\
                         &$\text{RMVPE}_\textit{Vocal}$  & \textbf{97.27±2.35} &\textbf{97.28±2.35} &\textbf{96.70±1.74} \\ \hline
\end{tabular}
}
\vspace{-2mm}
\end{table}

All results are shown in Table~\ref{tab:clean}. It can be observed that HARMOF0 achieves the highest performance on MDB-stem-synth and $\text{RMVPE}_\textit{Vocal}$ achieves the highest performance on MIR-1K. On MDB-stem-synth dataset, $\text{RMVPE}_\textit{Vocal}$ decreases only 0.83\% at RPA, 0.90\% at RCA, 0.79\% at OA compared with the best model HARMOF0. On MIR-1K dataset, $\text{RMVPE}_\textit{Vocal}$ outperforms the second best model HARMOF0 by 1.2\% at RPA, 0.7\% at RCA, 0.39\% at OA. Besides, $\text{RMVPE}_\textit{Poly}$ only decreases 0.56\% at RPA, 0.48\% at RCA compared with $\text{RMVPE}_\textit{Vocal}$, since $\text{RMVPE}_\textit{Poly}$ is trained on the mixture of MIR-1K, and the data distribution of mixture and clean vocals in MIR-1K is different. These results show that our model can also achieve a comparable performance with other models on monophonic music and clean vocals.

\section{Conclusion}\label{sec:conclusion}
In this paper, we propose a robust vocal pitch estimation model RMVPE to extract vocal pitches from polyphonic music. Our model adopts deep U-Net to directly extract effective hidden features so as to have the ability of handling both accompaniment and noise. 
Extensive experiments on MIR-1K, MIR\_ST500 and Cmedia datasets show that our model gets the best performance for vocal pitch estimation in polyphonic music. 
Moreover, rich experiments on MIR-1K with different types of noise over all levels show that RMVPE is robust for noise. In the future, we will explore the lighter model to achieve the same performance for vocal pitch estimation in polyphonic music.

\section{Acknowledgements}
This work is supported by the National Science Foundation of China under the grant 62272466, and Public Computing Cloud, Renmin University of China.
\bibliographystyle{IEEEtran}
\bibliography{template}

\begin{thebibliography}{10}
\providecommand{\url}[1]{#1}
\csname url@samestyle\endcsname
\providecommand{\newblock}{\relax}
\providecommand{\bibinfo}[2]{#2}
\providecommand{\BIBentrySTDinterwordspacing}{\spaceskip=0pt\relax}
\providecommand{\BIBentryALTinterwordstretchfactor}{4}
\providecommand{\BIBentryALTinterwordspacing}{\spaceskip=\fontdimen2\font plus
\BIBentryALTinterwordstretchfactor\fontdimen3\font minus
  \fontdimen4\font\relax}
\providecommand{\BIBforeignlanguage}[2]{{%
\expandafter\ifx\csname l@#1\endcsname\relax
\typeout{** WARNING: IEEEtran.bst: No hyphenation pattern has been}%
\typeout{** loaded for the language `#1'. Using the pattern for}%
\typeout{** the default language instead.}%
\else
\language=\csname l@#1\endcsname
\fi
#2}}
\providecommand{\BIBdecl}{\relax}
\BIBdecl

\bibitem{carroll2011fundamental}
J.~Carroll, S.~Tiaden, and F.-G. Zeng, ``Fundamental frequency is critical to
  speech perception in noise in combined acoustic and electric hearing,''
  \emph{The Journal of the Acoustical Society of America}, pp. 2054--2062,
  2011.

\bibitem{mir}
Y.~V.~S. Murthy and S.~G. Koolagudi, ``Content-based music information
  retrieval (cb-mir) and its applications toward the music industry: A
  review,'' \emph{ACM Computing Surveys (CSUR)}, 2018.

\bibitem{sa}
M.~K. Reddy and K.~S. Rao, ``Excitation modelling using epoch features for
  statistical parametric speech synthesis,'' \emph{Computer Speech \&
  Language}, 2020.

\bibitem{de2002yin}
A.~De~Cheveign{\'e} and H.~Kawahara, ``Yin, a fundamental frequency estimator
  for speech and music,'' in \emph{The Journal of the Acoustical Society of
  America}, 2002, pp. 1917--1930.

\bibitem{swipe}
A.~Camacho and J.~G. Harris, ``A sawtooth waveform inspired pitch estimator for
  speech and music,'' in \emph{The Journal of the Acoustical Society of
  America}, 2008, pp. 1638--1652.

\bibitem{mauch2014pyin}
M.~Mauch and S.~Dixon, ``pyin: A fundamental frequency estimator using
  probabilistic threshold distributions,'' in \emph{Proceeding of IEEE
  International Conference on Acoustics, Speech and Signal Processing
  (ICASSP)}, 2014, pp. 659--663.

\bibitem{kim2018crepe}
J.~W. Kim, J.~Salamon, P.~Li, and J.~P. Bello, ``Crepe: A convolutional
  representation for pitch estimation,'' in \emph{Proceeding of IEEE
  International Conference on Acoustics, Speech and Signal Processing
  (ICASSP)}, 2018.

\bibitem{deepf0}
S.~Singh, R.~Wang, and Y.~Qiu, ``Deepf0: End-to-end fundamental frequency
  estimation for music and speech signals,'' in \emph{Proceeding of IEEE
  International Conference on Acoustics, Speech and Signal Processing
  (ICASSP)}, 2021, pp. 61--65.

\bibitem{harmof0}
W.~Wei, P.~Li, Y.~Yu, and W.~Li, ``Harmof0: Logarithmic scale dilated
  convolution for pitch estimation,'' in \emph{IEEE International Conference on
  Multimedia and Expo (ICME)}, 2022, pp. 1--6.

\bibitem{open2019}
F.-R. St{\"o}ter, S.~Uhlich, A.~Liutkus, and Y.~Mitsufuji, ``Open-unmix-a
  reference implementation for music source separation,'' \emph{Journal of Open
  Source Software}, p. 1667, 2019.

\bibitem{demucs2019}
A.~D{\'e}fossez, N.~Usunier, L.~Bottou, and F.~Bach, ``Music source separation
  in the waveform domain,'' \emph{arXiv preprint arXiv:1911.13254}, 2019.

\bibitem{spleeter2020}
R.~Hennequin, A.~Khlif, F.~Voituret, and M.~Moussallam, ``Spleeter: a fast and
  efficient music source separation tool with pre-trained models,''
  \emph{Journal of Open Source Software}, p. 2154, 2020.

\bibitem{crn2019}
M.~Dong, J.~Wu, and J.~Luan, ``Vocal pitch extraction in polyphonic music using
  convolutional residual network.'' in \emph{Proceeding of INTERSPEECH}, 2019,
  pp. 2010--2014.

\bibitem{jdc2022}
S.~Kum, J.~Lee, K.~L. Kim, T.~Kim, and J.~Nam, ``Pseudo-label transfer from
  frame-level to note-level in a teacher-student framework for singing
  transcription from polyphonic music,'' in \emph{Proceeding of IEEE
  International Conference on Acoustics, Speech and Signal Processing
  (ICASSP)}, 2022, pp. 796--800.

\bibitem{bn}
S.~Ioffe and C.~Szegedy, ``Batch normalization: Accelerating deep network
  training by reducing internal covariate shift,'' in \emph{International
  Conference on International Conference on Machine Learning (ICML)}, 2015, p.
  448–456.

\bibitem{gru}
J.~Chung, {\c{C}}.~G{\"{u}}l{\c{c}}ehre, K.~Cho, and Y.~Bengio, ``Empirical
  evaluation of gated recurrent neural networks on sequence modeling,''
  \emph{arXiv preprint arXiv:1412.3555}, 2014.

\bibitem{salamon2017analysis}
J.~Salamon, R.~M. Bittner, J.~Bonada, J.~J. Bosch, E.~G{\'o}mez~Guti{\'e}rrez,
  and J.~P. Bello, ``An analysis/synthesis framework for automatic f0
  annotation of multitrack datasets,'' in \emph{Proceeding of International
  Society for Music Information Retrieval (ISMIR)}, 2017.

\bibitem{mir1k}
C.-L. Hsu and J.-S.~R. Jang, ``On the improvement of singing voice separation
  for monaural recordings using the mir-1k dataset,'' \emph{IEEE Transactions
  on Audio, Speech, and Language Processing (TASLP)}, pp. 310--319, 2010.

\bibitem{mirst500}
J.-Y. Wang and J.-S.~R. Jang, ``On the preparation and validation of a
  large-scale dataset of singing transcription,'' in \emph{Proceeding of IEEE
  International Conference on Acoustics, Speech and Signal Processing
  (ICASSP)}, 2021, pp. 276--280.

\bibitem{raffel2014mir_eval}
C.~Raffel, B.~McFee, E.~J. Humphrey, J.~Salamon, O.~Nieto, D.~Liang, D.~P.
  Ellis, and C.~C. Raffel, ``mir\_eval: A transparent implementation of common
  mir metrics,'' in \emph{Proceeding of International Society for Music
  Information Retrieval (ISMIR)}, 2014.

\bibitem{mcfee2015librosa}
B.~McFee, C.~Raffel, D.~Liang, D.~P. Ellis, M.~McVicar, E.~Battenberg, and
  O.~Nieto, ``librosa: Audio and music signal analysis in python,'' in
  \emph{Proceedings of the 14th python in science conference}, 2015, pp.
  18--25.

\bibitem{kingma2014adam}
D.~P. Kingma and J.~Ba, ``Adam: A method for stochastic optimization,''
  \emph{arXiv preprint arXiv:1412.6980}, 2014.

\bibitem{adt}
M.~Mauch and S.~Ewert, ``The audio degradation toolbox and its application to
  robustness evaluation,'' in \emph{Proceeding of International Society for
  Music Information Retrieval (ISMIR)}, 2013.

\bibitem{GIO2022}
X.~Sun, X.~Liang, Q.~He, B.~Zhu, and Z.~Ma, ``Gio: A timbre-informed approach
  for pitch tracking in highly noisy environments,'' in \emph{International
  Conference on Multimedia Retrieval (ICMR)}, 2022, pp. 480--488.

\end{thebibliography}

\end{document}